\documentclass[a4paper,fleqn,10pt]{article}
\pdfoutput=1
\usepackage{amsmath}
\usepackage{amssymb}
\usepackage{array}
\usepackage{calc}
\usepackage{longtable}
\usepackage{multirow}
\usepackage{wasysym}
\usepackage{graphicx}
\usepackage{xspace}
\usepackage{todonotes}

\numberwithin{equation}{section}
\usepackage{mciteplus}
\usepackage{
  pgf,
  tikz}
\usetikzlibrary{
  patterns,
  shapes.multipart,
  arrows,
  trees,
  scopes,
  decorations.pathreplacing,
  decorations.pathmorphing,
  decorations.markings,
  decorations.text,
  positioning,
  calc
}
\usepackage[a4paper,pdfborder={0 0 0}]{hyperref}
\usepackage[format=hang,labelfont=bf,hypcap=true]{caption}
\usepackage{subcaption}
\usepackage{sectsty}
\allsectionsfont{\sffamily}
\subsubsectionfont{\mdseries\itshape\large}
\makeatletter
\DeclareRobustCommand*{\bfseries}{%
  \not@math@alphabet\bfseries\mathbf
  \fontseries\bfdefault\selectfont
  \boldmath
}
\makeatother
\let\spreprint\empty
\newcommand{\preprint}[1]{\def\spreprint{\protect#1}}
\let\sinstitute\empty
\newcommand{\institute}[1]{\def\sinstitute{\protect#1}}
\makeatletter
\renewcommand{\maketitle}{\begingroup
  \null\thispagestyle{empty}%
    \ifx\spreprint\empty
      \vskip 5ex
    \else
      \flushright\large\spreprint\vskip 2ex
    \fi
    \vskip 5ex
    \flushleft
      {\sffamily\bfseries\huge\@title}\vskip 6ex
      \@author\vskip 2ex
      \ifx\sinstitute\empty
      \else
        {\small\sinstitute}
      \fi
    \vskip 5ex
  \endgroup
}
\makeatother
\renewenvironment{abstract}{\begin{center}
  {\large\sffamily\bfseries Abstract: }
  \begin{minipage}[t]{0.75\textwidth}
}{\end{minipage}\end{center}\vskip 10ex}

\numberwithin{equation}{section}
\newcommand{\Sherpa}{S\protect\scalebox{0.8}{HERPA}\xspace}
\newcommand{\dst}{\displaystyle}

\newcommand{\done}{{\rm d}}

\newcommand{\mc}[1]{\mathcal{#1}}

\hypersetup{
  pdfauthor={Stefan Hoeche, Silvan Kuttimalai, 
    Steffen Schumann, Frank Siegert}
  pdftitle={Beyond Standard Model calculations with Sherpa}
}
\preprint{SLAC-PUB 16170\\IPPP/14/105\\DCPT/14/210\\MCNET-14-35}
\author{Stefan H{\"o}che$^1$, Silvan Kuttimalai$^2$,
  Steffen Schumann$^3$, Frank Siegert$^4$}
\title{Beyond Standard Model calculations with Sherpa}
\institute{
  $^1$ SLAC National Accelerator Laboratory, 
  Menlo Park, CA 94025, USA\\  
  $^2$ Institute for Particle Physics Phenomenology,
  Durham University, Durham DH1 3LE, UK\\
  $^3$ II. Physikalisches Institut, Universit{\"a}t G{\"o}ttingen, 
  Friedrich-Hund-Platz 1, 37077 G{\"o}ttingen, Germany\\
  $^4$ Institut f{\"u}r Kern- und Teilchenphysik,
  TU Dresden, D--01062 Dresden, Germany\\}
\begin{document}
\maketitle
\begin{abstract}
We present a fully automated framework as part of the Sherpa event generator
for the computation of tree-level cross sections in beyond Standard Model scenarios,
making use of model information given in the Universal FeynRules Output format.
Elementary vertices are implemented into C++ code automatically and provided 
to the matrix-element generator Comix at runtime. Widths and branching ratios 
for unstable particles are computed from the same building blocks. 
The corresponding decays are simulated with spin correlations.
Parton showers, QED radiation and hadronization are added by Sherpa, 
providing a full simulation of arbitrary BSM processes at the hadron level.
\end{abstract}
\section{Introduction}
\label{sec:intro}

The quest for new-physics signals in collider data requires their detailed simulation. 
Comprehensive analyses of measurement sensitivities, exclusion limits or possibly 
anomalies often consider a variety of Beyond Standard Model (BSM) scenarios.
For each hypothesis, production cross sections need to be evaluated, and particle decay 
widths and branching ratios have to be computed. Realistic simulations further include 
spin correlations between production and decay. For simulations at the particle level, 
parton-shower effects and non-perturbative corrections must also be considered. 

Given the vast number of new-physics models, the automation of such
calculations is mandatory. In fact, in the past years enormous efforts were
made not only to automate leading-order calculations, but next-to-leading-order 
calculations as well. A variety of related tools have been constructed, ranging from 
Feynman rule generators like FeynRules~\cite{Christensen:2008py,*Alloul:2013bka} over
spectrum-generator generators like Sarah~\cite{Staub:2009bi,*Staub:2013tta}
to matrix-element generators like MadGraph~\cite{Alwall:2011uj},
MadGolem~\cite{Binoth:2011xi,*GoncalvesNetto:2012yt}, 
MadLoop~\cite{Hirschi:2011pa,*Alwall:2014hca}, 
Whizard~\cite{Kilian:2007gr} and Amegic~\cite{Krauss:2001iv} and 
particle-level event generators~\cite{Buckley:2011ms}, such as 
Herwig~\cite{Corcella:2000bw,*Bahr:2008pv},
Pythia~\cite{Sjostrand:2006za,*Sjostrand:2014zea}
and Sherpa~\cite{Gleisberg:2003xi,Gleisberg:2008ta}.
Each of them deals with particular aspects of the simulation. 
Specific protocols have been developed to guarantee consistent parameter
and event passing between the various tools~\cite{Skands:2003cj,*Alwall:2007mw,*Allanach:2008qq}.

In this paper we present the status and new developments regarding the 
simulation of new-physics signals with the event generator
\Sherpa~\cite{Gleisberg:2003xi,Gleisberg:2008ta}. Former versions of \Sherpa 
already supported quite a number of new-physics models. They were either 
built in as for example the MSSM~\cite{Hagiwara:2005wg}, the ADD 
model~\cite{Gleisberg:2003ue} and several others 
\cite{Dedes:2008bf,*Kilic:2008ub,*Schumann:2011ji}, or 
invoked through a dedicated interface to FeynRules~\cite{Christensen:2009jx}.
This interface was limited to vertices with color- and Lorentz-structures supported
by the matrix-element generator Amegic~\cite{Krauss:2001iv}.  
In the work presented here we lift these restrictions by extending 
the capabilities of \Sherpa's second built-in matrix-element
generator Comix~\cite{Gleisberg:2008fv} to account for almost arbitrary BSM 
scenarios. We generalize the recursive 
amplitude generation formalism to arbitrary $n$-point vertices, and we
automate the implementation of Lorentz calculators based on the model 
representation in the Universal FeynRules Output (UFO)~\cite{Degrande:2011ua}.
Part of our new generator is thus equivalent to {\sc Aloha}~\cite{deAquino:2011ub}.
At present we constrain ourselves to particles of spin-0, spin-1/2 and spin-1.
A generalization to spin-3/2 and spin-2 states is straight-forward and foreseen 
for the near future. Similarly, we restrict ourselves to color structures
involving singlets, (anti-)triplets, and octets. (Anti-)sextet representations
will be included in the near future. We also discuss the implementation of an algorithm
to preserve spin correlations between factorized production and decay
processes~\cite{Richardson:2001df}.

This paper is organized as follows. In Sec.~\ref{sec:amplitudes}
we discuss the techniques used for amplitude generation focusing on
the newly developed methods for the automatic implementation of Lorentz
structures. We also present the results of an extensive validation.
In Sec.~\ref{sec:decays} we introduce
and discuss our treatment of particle decays, including spin-correlation 
effects. After a discussion of other event generation aspects in 
Sec.~\ref{sec:embedding} the conclusions and an outlook 
are given in Sec.~\ref{sec:conclusions}.

\section{Cross-section calculations at tree-level}
\label{sec:amplitudes}
This section briefly describes the algorithms implemented in the matrix-element generator
Comix to compute tree-level amplitudes. Identical methods are used to obtain tree-level 
like objects for next-to-leading order calculations, i.e.\ the color-correlated 
Born amplitudes entering dipole-subtraction terms in the Catani--Seymour 
method~\cite{Catani:1996vz,*Catani:2002hc} or the FKS method~\cite{Frixione:1995ms}.
The implementation of dipole-subtraction in Comix will be described 
elsewhere~\cite{Hoeche:2014xx}.

A recursive algorithm for the computation of color-ordered multi-parton amplitudes 
was proposed long ago~\cite{Berends:1987cv,*Berends:1987me}. Its extension to colorful
amplitudes~\cite{Duhr:2006iq} leads to a recursion that resembles the Dyson--Schwinger
equations~\cite{Dyson:1949ha,*Schwinger:1951ex,*Schwinger:1951hq}. In this publication
we extend the implementation of the algorithm in the matrix-element generator 
Comix~\cite{Gleisberg:2008fv} such that it can handle $n$-point vertices at tree level,
where $n$ is -- in principle -- unbounded. The automatic implementation of related 
Lorentz structures is described in Sec.~\ref{sec:lorentz}.

\begin{figure}[t]
  \begin{center}
    \raisebox{-4.35mm}{\includegraphics[scale=0.4]{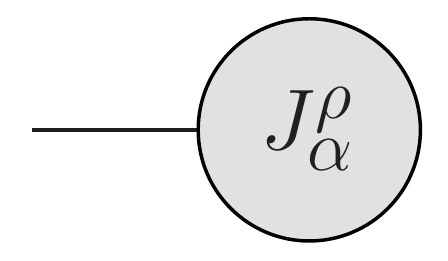}}
    $\dst=\!\!\sum\limits_{\substack{\{\rho_1,\rho_2\}\\\in O\!P_2(\rho)\\[1mm]V_\alpha^{\alpha_1\alpha_2}}}\!\!$
    \raisebox{-13.3mm}{\includegraphics[scale=0.4]{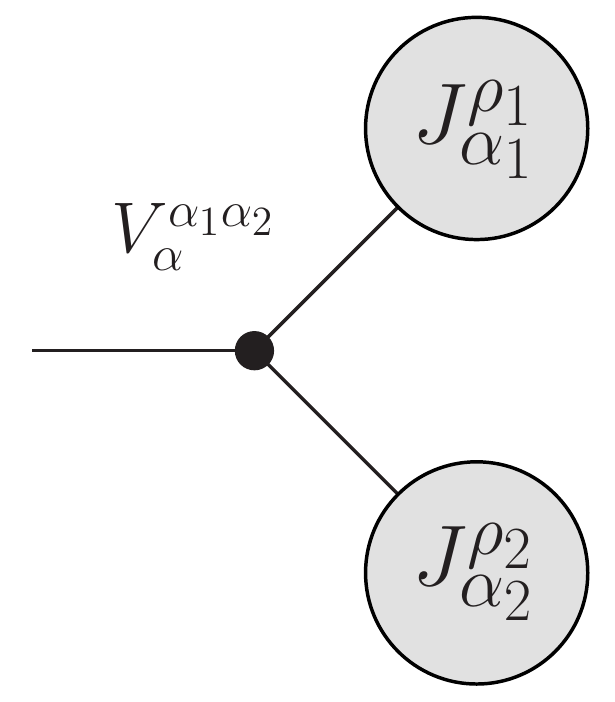}}
    $\dst+\!\!\sum\limits_{\substack{\{\rho_1,\rho_2,\rho_3\}\\\in O\!P_3(\rho)\\[1mm]V_\alpha^{\alpha_1\alpha_2\alpha_3}}}\!\!$
    \raisebox{-15.6mm}{\includegraphics[scale=0.4]{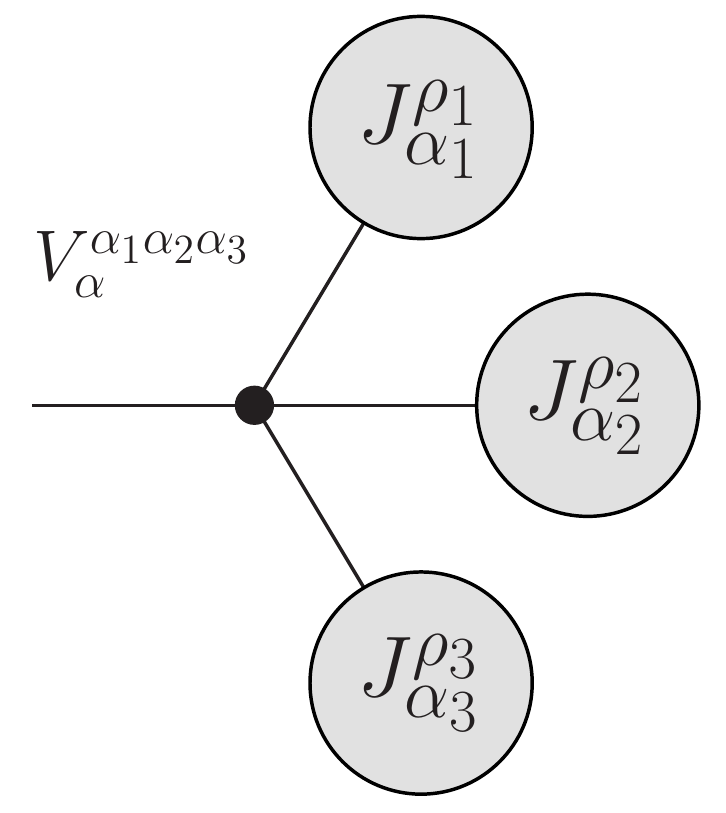}}
    $\dst+\;\ldots\;+\!\!\sum\limits_{\substack{\{\rho_1,\rho_2,\ldots,\,\rho_n\}\\\in O\!P_n(\rho)\\[1mm]V_\alpha^{\alpha_1\alpha_2\ldots\,\alpha_n}}}\!\!$
    \raisebox{-18.4mm}{\includegraphics[scale=0.4]{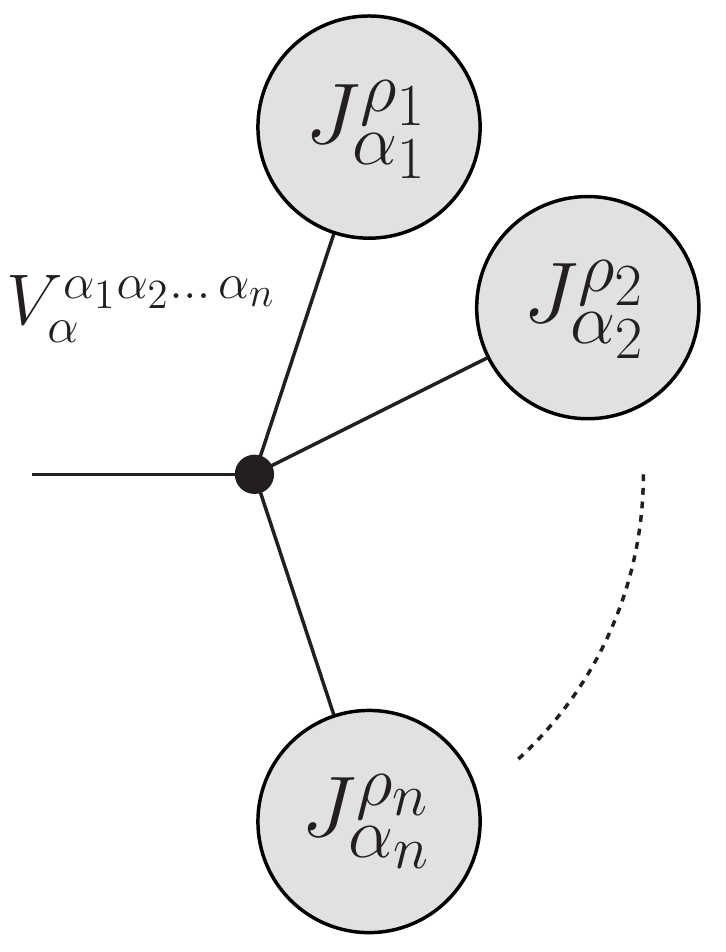}}
  \end{center}
  \caption{Sketch of the Berends--Giele type recursive relation as implemented in Comix.
    The current $J^\rho_\alpha$ is computed as a sum over sub-currents joined by elementary
    vertices. This formulation is inherently recursive. The sums on the right-hand side extend
    over all ordered partitions of the set of particles, $\rho$, on the left hand side.
    The multi-indices $\alpha$ denote both Lorentz and color indices of the currents.
    Displayed are vertices with up to $n+1$ external particles.
    \label{fig:bg_recursion}}
\end{figure}
Schematically the algorithm to compute tree-level amplitudes based on the Berends--Giele
type recursive relations is depicted in Fig.~\ref{fig:bg_recursion}.
Consider an unordered $N$-particle current, $J^\rho_\alpha$, where $\rho$ denotes
the set of $N$ particles, and $\alpha$ is a multi-index that labels both Lorentz
and color indices of the current. This current is computed from all Feynman graphs
having as external particles the on-shell particles in the set $\rho$, and the
(potentially off-shell) particle described by $J^\rho_\alpha$. Special currents
are given by the external-particle currents. They correspond to the helicity Eigenvectors
of wave functions for the external particles, as described in~\cite{Gleisberg:2008fv}.
Assuming that up to $n+1$-point vertices exist, off-shell currents can be computed as
\begin{equation}\label{eq:bg_recursion}
  J^\rho_\alpha\;=\;P^{\,\rho}_{\alpha}\;
  \sum_{m=2}^n\;
  \sum\limits_{\substack{\{\rho_1,\ldots,\,\rho_m\}\\\in\, O\!P_m(\rho)}}\;
  \sum\limits_{V_{\alpha}^{\,\alpha_1\ldots\,\alpha_m}}\;S(\rho_1,\ldots,\,\rho_m)\;
  V_{\alpha}^{\,\alpha_1\ldots\,\alpha_m}\;J^{\rho_1}_{\alpha_1}\,\ldots\,J^{\rho_m}_{\alpha_m}\;.
\end{equation}
Here $P^\rho_\alpha$ denotes a propagator term depending on the particle type $\alpha$
and the set of particles $\rho$. The sum over $V$ extends over all elementary vertices
of the theory that have as external states the particles described by the currents
$J^{\rho_1}_{\alpha_1}\ldots J^{\rho_m}_{\alpha_m}$. For some assignment of currents
no such vertex may exist. The final sum extends over all ordered partitions of the
set of indices in $\rho$. $S$ is the symmetry factor associated with the decomposition
of $\rho$ into subsets, see \cite{Gleisberg:2008fv}.

An $N$-particle scattering amplitude is given in terms of the above current as
\begin{equation}\label{eq:bg_amplitude}
  \mc{A}(1,\ldots,N)\,=\;J^{\{N\}}_\alpha\,
    \frac{1}{P^{\{1,\ldots,N-1\}}_\alpha}\,
    J^{\,\{1,\ldots,N-1\}}_\alpha\;.
\end{equation}
Note that the amputation of the final propagator term is schematic. In practice,
one does not multiply with this term in the first place.

In order to implement Eqs.~\eqref{eq:bg_recursion} and \eqref{eq:bg_amplitude}
we employ the spinor basis introduced in Ref.~\cite{Hagiwara:1985yu}.
The $\gamma$-matrices are taken in the Weyl representation, which has the advantage
that massless spinors are described by only two nonzero components. 
Polarization vectors for external vector bosons are constructed according to
Ref.~\cite{Dittmaier:1998nn}.

Majorana fermions are treated in the formalism of~\cite{Denner:1992vza,*Denner:1992me}.
Their external wave functions can be constructed either as if they represent fermions,
or as if they represent anti-fermions. This is left optional in Comix, and it can be 
used to check the consistency of the calculation.

Comix allows to specify coupling orders for the calculation. This
permits, for example, to compute only strongly interacting parts of
$pp\to jj$ amplitudes, or exclusively electroweak contributions. In
the UFO format, not only the QCD and electroweak order of a coupling
can be specified. Instead, arbitrary orders can be defined and the
coupling constants are classified accordingly. This feature is fully
supported and by default no restrictions with respect to coupling
orders are applied. If instead the user specifies a coupling
constraint, Comix applies this constraint at the amplitude-squared
level. It is therefore also possible to compute pure interference terms.
While these terms are not observable in practice, computing them is
often instructive to study directly the difference between coherent
and incoherent sums of signal and background contributions.

\subsection{Treatment of color}
\label{sec:colour}

Comix samples external colors and performs the color algebra in the color-flow 
decomposition at the vertex level. The color-flow decomposition, formally introduced 
in~\cite{tHooft:1973jz}, was advertised in the context of collider physics 
in~\cite{Maltoni:2002mq}. It was shown to be superior for high-multiplicity
QCD calculations in~\cite{Duhr:2006iq}.

In the color-flow decomposition, each particle in the adjoint representation is replaced
by a bi-fundamental, while keeping track of the active degrees of freedom by applying
projection operators. This amounts to cutting adjoint propagators by inserting the 
identity $\delta^{ab}=T^a_{ij}T^b_{ji}$ and identifying $i$ and $j$ as the propagator
indices. In practice one contracts adjoints with generators at vertices, while 
inserting projectors of the form $T^a_{ij}T^a_{kl}$ in each propagator. 

We have implemented the relevant color structures for the Standard Model, the MSSM, and
a range of BSM theories. This includes the trivial identities, group generators, structure 
constants as well as simple products of those. Color (anti-)sextets can be accomodated,
but our code does not include them at present. The implementation of Standard Model color
structures has been detailed in~\cite{Duhr:2006iq}. It is straightforward to implement 
higher-point functions, and the corresponding objects can be supplied to Comix
at runtime using a dynamically linked library. So far we have not automated the generation
of color calculators, but there is no obstacle to do so.

\subsection{Automatic implementation of Lorentz calculators}
\label{sec:lorentz}

\begin{figure}[t]
  \begin{center}
    \tikzset{
  outg/.style=
  {
    densely dashed,
    thick,
    draw=black,
  },
  inc/.style=
  {
    draw=black
  },
  vert/.style=
  {
    on grid, 
    thick, 
    draw=black, 
    circle, 
    inner sep=2.0pt, 
    outer  sep=0pt
  }
}

\pgfmathsetmacro{\r}  {1.2}
\pgfmathsetmacro{\fac}{0.9}

\begin{tikzpicture}[]
  \node [vert]    (z)     at (0.00,0.00)      {$\Gamma$};
  \node [on grid] (j0)    at ($(z)+(-180:\fac*\r)$) {$\alpha_0$};
  \node [on grid] (j1)    at ($(z)+(45:\r)$) {$J_1$};
  \node [on grid] (j2)    at ($(z)+(20:\r)$) {$J_2$};
  \node [on grid] (jn)    at ($(z)+(-45:\r)$) {$J_n$};
  \coordinate          (start) at ($(z)+( 0:\fac*\r)$);
  \draw [dotted,thick] (start) arc[radius =\fac*\r, start angle=0, end angle=-30];
  \path [outg]    (z)   to    (j0);
  \path [inc]     (z)   to    (j1);
  \path [inc]     (z)   to    (j2);
  \path [inc]     (z)   to    (jn);
  \node [on grid] (eq)    at ($(z) - (0,1.7*\r)$) {$\Gamma_{\alpha_0}^{\;\alpha_1\ldots\,\alpha_n}J_{\alpha_1}\ldots J_{\alpha_n}$};

  \put (175,0) {$\ldots$};
  
  \begin{scope}[shift={(4,0)}]
    \node [vert]    (z)     at (0.00,0.00)      {$\bar\Gamma$};
    \node [on grid] (j0)    at ($(z)+(-180:\fac*\r)$) {$\alpha_1$};
    \node [on grid] (j1)    at ($(z)+(45:\r)$) {$J_2$};
    \node [on grid] (j2)    at ($(z)+(20:\r)$) {$J_3$};
    \node [on grid] (jn)    at ($(z)+(-45:\r)$) {$J_0$};
    \coordinate          (start) at ($(z)+( 0:\fac*\r)$);
    \draw [dotted,thick] (start) arc[radius =\fac*\r, start angle=0, end angle=-30];
    \path [outg]    (z)   to    (j0);
    \path [inc]     (z)   to    (j1);
    \path [inc]     (z)   to    (j2);
    \path [inc]     (z)   to    (jn);
    \node [on grid] (eq)    at ($(z) - (0,1.7*\r)$) {$\bar\Gamma_{\alpha_1}^{\;\alpha_2\ldots\,\alpha_0}J_{\alpha_2}\ldots J_{\alpha_0}$};
  \end{scope}

  \begin{scope}[shift={(9,0)}]
    \node [vert]    (z)     at (0.00,0.00)      {$\tilde\Gamma$};
    \node [on grid] (j0)    at ($(z)+(-180:\fac*\r)$) {$\alpha_n$};
    \node [on grid] (j1)    at ($(z)+(45:\r)$) {$J_0$};
    \node [on grid] (j2)    at ($(z)+(20:\r)$) {$J_1$};
    \node [on grid] (jn)    at ($(z)+(-45:\r)$) {$J_{n-1}$};
    \coordinate          (start) at ($(z)+( 0:\fac*\r)$);
    \draw [dotted,thick] (start) arc[radius =\fac*\r, start angle=0, end angle=-30];
    \path [outg]    (z)   to    (j0);
    \path [inc]     (z)   to    (j1);
    \path [inc]     (z)   to    (j2);
    \path [inc]     (z)   to    (jn);
    \node [on grid] (eq)    at ($(z) - (0,1.7*\r)$) {$\tilde\Gamma_{\alpha_n}^{\;\alpha_0\ldots\,\alpha_{n-1}}J_{\alpha_0}\ldots J_{\alpha_{n-1}}$};
  \end{scope}

\end{tikzpicture}

  \end{center}
  \caption{Pictorial representation of the possible rotations of Lorentz calculators.
    \label{fig:lc}}
\end{figure}
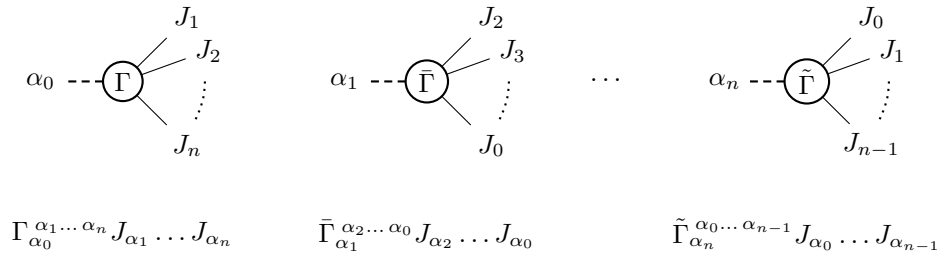
Within the Dyson-Schwinger formalism discussed above, any
off-shell current corresponds to a particle and therefore one specific
quantum field and its representation of the Lorentz group. Although
the implementation in Comix is currently limited to spin-0, spin-1/2,
and spin-1 particles (including the spin-2 pseudoparticle described in~\cite{Duhr:2006iq}), 
our automatic implementation of numerical routines for evaluating the Lorentz structures 
of vertices is generic.
It only requires, that currents be represented by multi-component complex
objects and that the recursive relations, Eq.~\eqref{eq:bg_recursion}, are used.
For each model, routines must be provided that evaluate expressions of the form
$\Gamma_{\alpha_0}^{\;\alpha_1\ldots\,\alpha_n}J_{\alpha_1}\ldots J_{\alpha_n}$,
which correspond to the space-time structure of the vertices in
Eq.~\eqref{eq:bg_recursion}. Pictorially, one can represent such terms
as shown in figure~\ref{fig:lc}. 

The Universal FeynRules Output (UFO)~\cite{Degrande:2011ua} is a format for 
exchanging information on interaction vertices in terms of a basic set of 
color and Lorentz structures and symbolic algebraic operations on those. 
We have constructed a Python module that implements explicit representations
of the Lorentz structures as they are used in Comix and maps them
onto the definitions in the UFO. This module is capable of performing all
algebraic operations on these building blocks to generate C++ source
code to be used by Comix for the corresponding Lorentz calculators.

With the UFO expression for an $n+1$-particle vertex at hand, the Python module 
sets up external currents $J_{\alpha_1},\ldots,J_{\alpha_n}$ with symbolic components
and then performs the multiplications and implicit sums over indices, leaving only 
the ``outgoing'' index, $\alpha_0$, uncontracted.
This yields an explicit expression for all components of the current $J_{\alpha_0}$ 
that is stored in the form of C++ code. Note that this procedure needs to be
performed for all cyclic permutations of indices $\{0,\ldots,n\}$,
each one corresponding to a different ``outgoing'' index. Pictorially,
this corresponds to a counter-clockwise rotation of the vertex, as
shown in Fig.~\ref{fig:lc}.

As an example, consider the gauge coupling of a vector field $A^\mu$ to a 
fermion, $\overline{\psi}\gamma_\mu A^\mu\psi$. Taking $\alpha_0$
to be the vector, and $\alpha_1$ and $\alpha_2$ to be the Dirac antiparticle 
and particle, respectively, the Lorentz calculator schematically depicted
in \ref{fig:lc} would correspond to
\begin{align}
(\gamma^\mu)_{ab}\overline{\psi}_a\psi_b\quad=\;
\begin{aligned}\tikzset{
  outg/.style=
  {
    densely dashed,
    thick,
    draw=black,
  },
  inc/.style=
  {
    draw=black
  },
  vert/.style=
  {
    on grid, 
    thick, 
    draw=black, 
    circle, 
    inner sep=1pt, 
    outer  sep=0pt
  }
}
\pgfmathsetmacro{\r}  {1.2}
\pgfmathsetmacro{\fac}{0.9}
\begin{tikzpicture}[scale=1]
  \node [vert]    (z)     at (0.00,0.00)      {$\gamma^\mu\vphantom{\int}$};
  \node [on grid] (j0)    at ($(z)+(-180:\fac*\r)$) {$\mu$};
  \node [on grid] (j1)    at ($(z)+(45:\r)$) {$\overline{\psi}$};
  \node [on grid] (jn)    at ($(z)+(-45:\r)$) {$\psi$};
  \path [outg]    (z)   to    (j0);
  \path [inc]     (z)   to    (j1);
  \path [inc]     (z)   to    (jn);
\end{tikzpicture}
\end{aligned}\;.
\end{align}
Analogous expressions must be provided for the other two cyclic permutations 
of the indices $\{0,1,2\}$.

\subsection{Implementation of model parameters}
\label{sec:model_impl}
The C++ routines generated in this manner are compiled and linked along with the 
information on the particle content of the model and the model parameters.
The dynamic library containing Lorentz calculators and model information 
is loaded by \Sherpa at program startup. The entire process is automated to a high level,
such that the user needs to run just a single command to make the entire UFO model
available for event generation.

The parameters of the model are set to the default values given in the UFO. 
They can be overwritten at runtime using a file which largely follows the 
SLHA~\cite{Skands:2003cj,*Alwall:2007mw,*Allanach:2008qq}.
Note that at this level it is not possible anymore to change parameters which
would lead to the appearance
of additional vertices in the model, like changing the Yukawa mass of a bottom
quark from zero to a nonzero value. The set of model parameters is available 
throughout the whole \Sherpa framework, which guarantees the consistent use of couplings 
and particle masses at all stages of event generation.

\subsection{Illustrative examples}
\label{sec:me_examples}
\begin{figure}[t]
  \centering
  \includegraphics[width=0.55\textwidth]{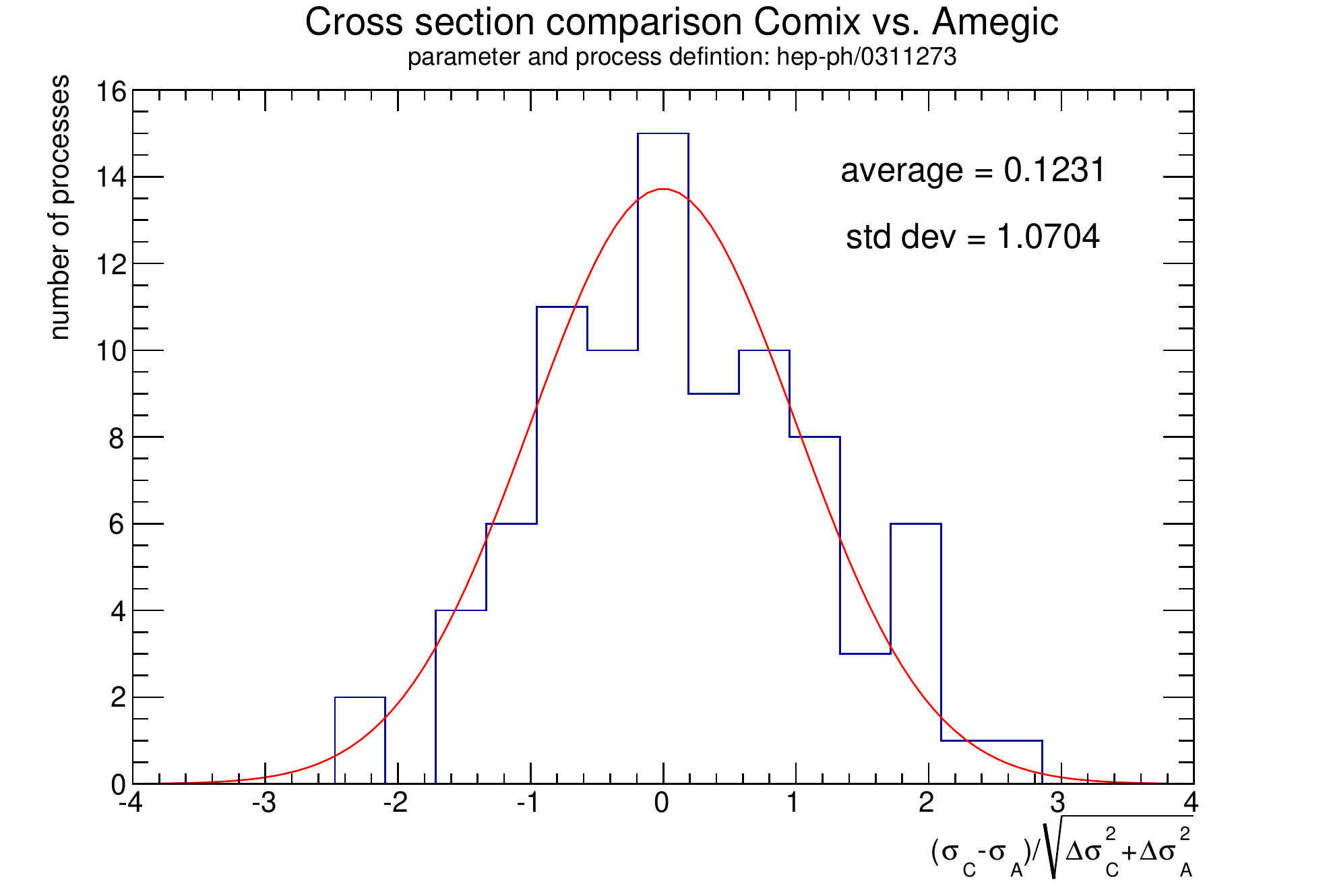}
  \caption{Deviation between results from Amegic and Comix for the 86
    $e^+e^-\to 6f$ processes listed in Ref.~\cite{Gleisberg:2003bi},
    using the parameters given ibidem. The red curve represents a normal
    distribution and should be considered the reference. 
    \label{fig:ilc_test}}
\end{figure}

\begin{table}[t]
  \centering
  \begin{tabular}{l@{\qquad\qquad}cc}
    Model & number of & max. rel. deviation\\
          & processes tested & Comix $\leftrightarrow$ MadGraph5 \\\hline\\[-2mm]
    Standard Model                     & 60  & $2.3\cdot 10^{-10}$  \\ %
    Higgs Effective Field Theory       & 13  & $4.3\cdot 10^{-13}$  \\ 
    MSSM                               & 401 & $1.0\cdot 10^{-10}$  \\ 
    Minimal Universal Extra Dimensions & 51  & $2.8 \cdot 10^{-12}$ \\ 
    Anomalous Quartic Gauge Couplings  & 16  & $5.9\cdot 10^{-12}$  \\[2mm] 
    \hline
  \end{tabular}
  \caption{Maximal relative deviations between tree-level matrix elements
    computed with Comix and MadGraph5. For each model we quote the largest 
    observed deviation among all processes, where we tested 1000 random 
    phase-space points per process.
    \label{tab:bsm_test}}
\end{table}

In order to validate our new generator we compared numerous results 
obtained with Comix for a variety of models against Amegic~\cite{Krauss:2001iv} 
and MadGraph5~\cite{Alwall:2011uj}.

Figure~\ref{fig:ilc_test} shows the deviation of leading-order cross sections 
computed both with Amegic and Comix for the 86 $e^+e^-$ to six-fermion processes 
listed in Ref.~\cite{Gleisberg:2003bi}, where each result is 
computed to better than $5\permil$ Monte-Carlo uncertainty. It can be seen that the 
deviation between the two generators is of purely statistical nature. This confirms 
the correct implementation of the Standard Model in the extended version of Comix,
and it validates the recursive phase-space generator described in~\cite{Gleisberg:2008fv}.

Table~\ref{tab:bsm_test} presents a comparison of tree-level matrix
elements between Comix and MadGraph5. In all tests we have considered
1000 individual phase-space points per process. For each model we
quote only the maximal deviation found when comparing matrix elements
from MadGraph5 and Comix. We considered the processes and parameters
listed in~\cite{Christensen:2009jx} for the Minimal Universal Extra
Dimensions Model. In case of the MSSM, we tested the more
comprehensive set of processes considered in~\cite{Hagiwara:2005wg}
and the set of processes considered in~\cite{Christensen:2009jx} for
the Standard Model was also supplemented by further $2\rightarrow 2$,
$2\rightarrow 3$, and $2\rightarrow 4$ processes.

We have also compared the results from Comix against those from
MadGraph5 for two effective theories. The first is based on the 
Standard Model including couplings of a scalar and a hypothetical 
pseudoscalar Higgs boson to gluons via a top-quark
loop~\cite{Ellis:1975ap,*Wilczek:1977zn,*Shifman:1979eb,*Ellis:1979jy}.
This theory involves up to five-point vertices. In order to test our
algorithms in the context of more complicated Lorentz structures and
high-multiplicity vertices, we considered anomalous quartic gauge
couplings~\cite{Belyaev:1998ih,*Eboli:2000ad,*Eboli:2003nq}.
Specifically, we used a model implementing the interaction terms (A7)
- (A10), as described in~\cite{Eboli:2006wa}. They give rise to up to
eight-particle vertices extending the gauge sector of the Standard
Model. We tested $2\rightarrow 2$ as well as $2\rightarrow 4$
processes that are sensitive to complicated Lorentz structures of
up to 6-particle vertices which cannot be mapped to Standard-Model 
like interactions. The number of processes compared and the maximal
relative deviation observed are again listed in Table~\ref{tab:bsm_test}.
This successful validation proves that effective operators can 
efficiently be implemented in Comix via FeynRules and UFO.

\section{Decay simulation including spin correlations}
\label{sec:decays}

It is often not feasible to simulate new-physics signals at the
level of stable final-state particles. The possibility of many intermediate 
resonances leads to a large number of different final states. Even if 
matrix-element calculation and phase-space integration for each of those 
final states are in principle feasible, the management of all possible states 
within a matrix-element generator becomes computationally challenging and
practically useless. It is more convenient to simulate only the production
of certain new-physics resonances, and possibly the accompanying hard QCD
and/or QED radiation, while treating the cascade decay of heavy unstable
new-physics objects in a different manner.

Here we describe a module of the \Sherpa event generator which implements such
a decay cascade. It performs two main tasks which will be described in
the next subsections:
the construction of the cascade itself, and
the preservation of spin correlations which are neglected during the
independent calculation of production and decay in the cascade.

\subsection{Construction of the decay cascade}
\label{sec:decays:casc}
To construct a decay cascade one recursively simulates single decay processes
until only stable particles are left. For the simulation of each single decay
process several ingredients are necessary.

The first step is the choice of a decay channel according to its branching
ratio.
The basic information for determining possible decay modes of a given unstable
particle $P$ are the vertices, $V$, of Eq.~\eqref{eq:bg_recursion},
which contain $P$ among their $n$ external lines.
Using these vertices as a starting point, an initial (direct) decay table is
built up for potential $P\to n-1$ decay modes.

Each decay mode can then be revisited to decide whether it is accepted as final
or whether it should be replaced by including further iterative
decays\footnote{This implies that the decay tables are initialized in the order of
  the unstable particle masses.}.
The simplest option for this decision is the mass threshold criterion: if
the mass of the outgoing system is larger than the decayer mass, then the
direct decay mode is discarded and replaced by all possible combinations where
one final-state particle has been replaced by its own decay products. When a
decay mode is replaced, only diagrams with the given propagator structure
should be included in the matrix elements for the new decay channels.
For cases where the threshold criterion is too simple an alternative option is
implemented where the decision is triggered by a comparison of the partial
widths calculated from the direct vs.\ the converted decay modes. If more
sophisticated threshold behavior is necessary the user of \Sherpa can implement 
a dedicated trigger criterion involving e.g.\ additional phase-space weights.
This conversion of decay modes could be iterated. In our implementation we allow
for one step, which should be sufficient for most practical applications. 
Assuming e.g.\ only 3-point vertices for simplicity this allows for a conversion 
from $1\to 2$ modes to $1\to 3$ modes.
Depending on the complexity of the model it can take a few minutes to construct
the decay table. Considering for example the MSSM model with the SPS1a
benchmark point~\cite{AguilarSaavedra:2005pw}, we find that the construction of
the decay table takes 150 seconds using one core of an Intel Xeon E5-2670 CPU
at 2.6GHz and requires 0.7 GB of main memory.
To facilitate a quick initialization for the case of more complex models it is
possible to write the decay table to disk and read it back in.

For each final decay channel the corresponding matrix element is
constructed using the building blocks described in Sec.~\ref{sec:amplitudes}.
This implies that the full BSM capabilities stemming from the UFO implementation
are available also in the decay module.
We consider tree-level amplitudes only, using the exact same
model parameters as for the hard-scattering process, cf. Sec.~\ref{sec:model_impl}.
Integrating a decay matrix element over phase space one obtains
the partial width of that channel and correspondingly its selection probability
in the decay table.

These matrix elements are also used to go beyond an isotropic distribution of
the decay kinematics. For simple two-body decays, the phase space is generated
using the Rambo algorithm~\cite{Kleiss:1985gy}. For decays to three and more
particles we employ importance-sampling based on information about
propagators~\cite{Byckling:1969sx}. If applicable several channels are combined
into a multi-channel integrator~\cite{Kleiss:1994qy}. The matrix elements are
then used in an unweighting step to provide the final decay kinematics.

The full amplitude-level 
information including the helicity dependence is also made available to allow 
for the implementation of spin correlations, as will be described in the
following section.

As an additional option to improve the modeling of decay cascades we
implement a crude estimation of off-shell effects by adjusting the decay
kinematics a posteriori to yield a Breit--Wigner distribution of the decayer
momentum. This is at the present based on a constant-width approach and can 
in the future be improved with dedicated line-shape modeling in selected cases.

\subsection{Spin-correlation algorithm}
\label{sec:decays:sc}
The factorization into production and decay matrix elements is based on the
replacement of intermediate particle propagators by a helicity sum, using completeness relations.
For example, a full matrix element containing a massive vector-boson propagator
can be factorized as:
\begin{equation}
  \label{eq:decays:fact}
  \mathcal{M} \;\sim\;
  j_1^\mu \,\left[\,g_{\mu\nu}-\frac{p_\mu p_\nu}{p^2}\,\right]\,j_2^\nu
  \;=\;
  \sum\limits_{\lambda} 
  \underbrace{j_1^\mu\varepsilon_\mu^*(\lambda)}_{\mathcal{M}_\text{prod}(\lambda)} \;
  \underbrace{\varepsilon_{\nu\vphantom{_\mu}}(\lambda) j_2^\nu}_{\mathcal{M}_\text{dec}(\lambda)}\;.
\end{equation}
Similar equations hold for all particle types.
If spin correlations were neglected, the common sum over helicities $\lambda$ 
would be replaced by individual sums for production and decay.
While for some applications this is a reasonable approximation, in other cases
it will lead to a significant mis-modeling, e.g.\ of angular correlations
between decay products. To remedy this situation we
employ a spin-correlation algorithm originally introduced for QCD parton
showers~\cite{Collins:1987cp,Knowles:1987cu,*Knowles:1988vs} and generalized
to arbitrary decay cascades in~\cite{Richardson:2001df}.

In this algorithm, the helicity summation or averaging in a
matrix element is replaced at each step by a contraction with the
spin-density matrices $\rho_{\lambda\lambda'}$ of the incoming particles
and the decay matrices $D_{\lambda\lambda'}$ of unstable outgoing particles:
\begin{equation}
  \label{eq:decays:scdgamma}
  \frac{\done\Gamma(0\to 1\dots n)}{\done\Omega} = \rho_{\lambda_0\lambda'_0}\;
  \mathcal{M}_{\lambda_0;\lambda_1,\dots\lambda_n}\mathcal{M}^*_{\lambda'_0;\lambda'_1,\dots\lambda'_n}\;
  \prod_{i=1,n} D^i_{\lambda_i\lambda'_i}\;.
\end{equation}
These are not fully known at all stages of the decay cascade though, 
and~\cite{Richardson:2001df} describes the algorithm with which they can be
continually updated and implemented as they become available.

We obtain the full helicity structure of the amplitudes
$\mathcal{M}_{\lambda_0;\lambda_1,\dots\lambda_n}$
from our decay matrix-element generator described in Sec.~\ref{sec:decays:casc}.
We use the same building blocks and gauge conventions in the production
and decay matrix elements, therefore the algorithm will directly recover
the spin correlations in the decay cascade.

To demonstrate these features, we are presenting one example in the Standard
Model, namely top-quark pair production, and one in the MSSM, namely the production 
of a squark pair with subsequent decay cascades.

\subsubsection{Top-quark pair production in the SM}
\label{sec:decay_results_sm}
For our simulation of top-quark production at the LHC we consider exclusively 
the decay $t\to Wb$, while the resulting $W$-boson pair decays into an electron 
and muon final state according to $W^+ \to e^+ \nu_e$ and $W^-\to \mu^- \bar{\nu}_\mu$.
We present results for three different approaches to simulate this final state:
\begin{description}
\item[Full ME] \hfill\\
  The full $pp\to e^+ \nu_e \mu^- \bar{\nu}_\mu b \bar{b}$ final state is
  simulated in the Comix 
  matrix-element generator, with a restriction to doubly-resonant diagrams and
  onshell intermediate top quarks and $W$ bosons. This automatically includes
  all helicity correlations by construction and is thus used as a reference.
\item[Correlated decays] \hfill\\
  Only the $pp\to t\bar{t}$ process is generated as
  hard scattering with the Comix generator. The decays are simulated in a
  factorized manner and spin correlations are implemented as described above.
\item[Uncorrelated decays] \hfill\\
  As above, but without implementing spin-correlations.
\end{description}

\begin{figure}
  \centering
  \begin{minipage}[t]{0.5\linewidth}
    \includegraphics[width=\textwidth]{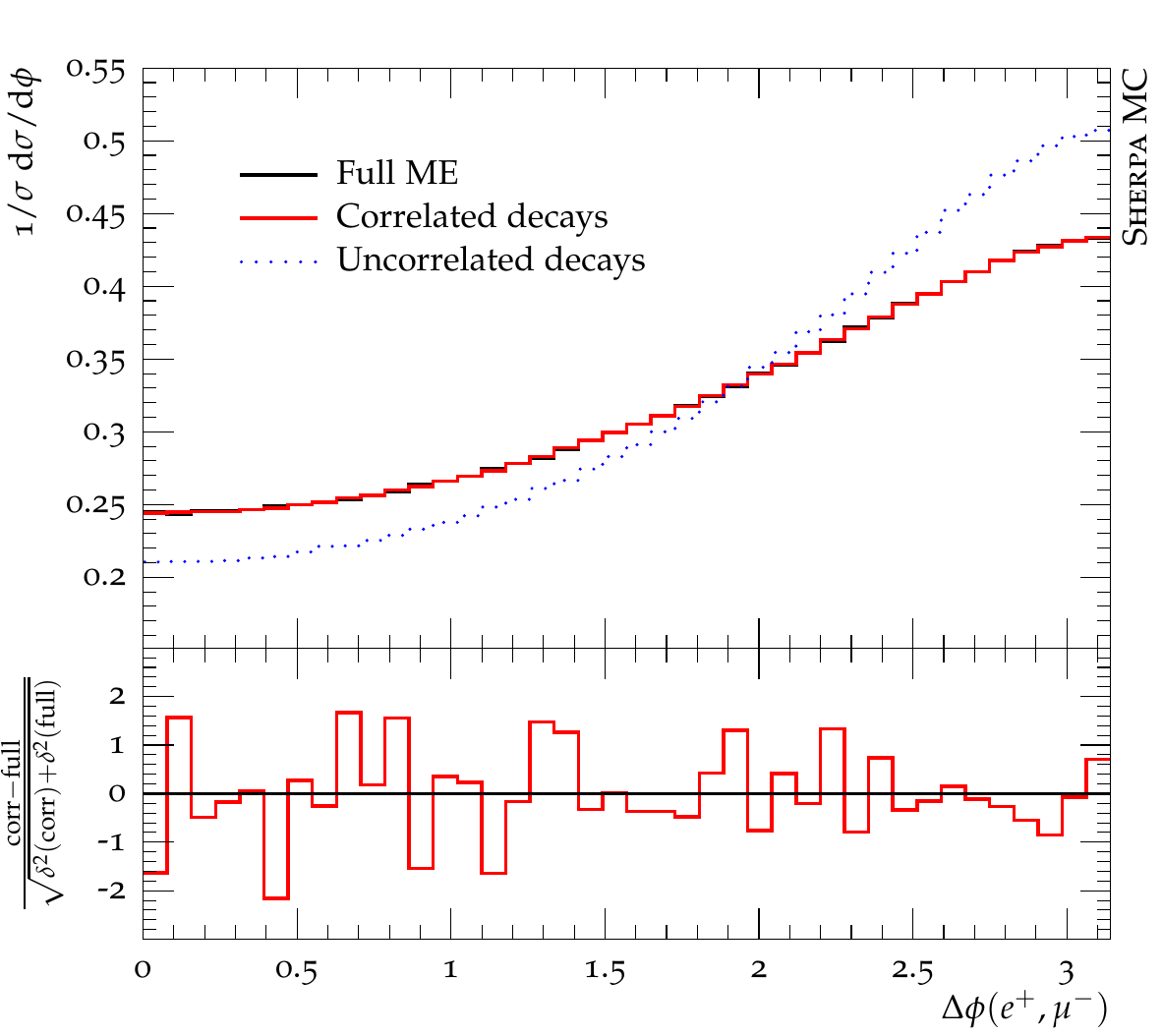}
  \end{minipage}\nolinebreak
  \caption{Spin-correlation effects in top-quark pair production in the SM.
    The three simulation setups are described in the text. The ratio plot
    displays the relative difference in terms of the standard deviation and
    allows to judge the statistical compatibility between the full ME and
    correlated decay simulation.
  }
  \label{fig:decays:sc1}
\end{figure}

In Fig.~\ref{fig:decays:sc1} we present the comparison of the three different
approaches for the azimuthal separation of the muon and positron. It can
already be seen in this simple example that the simulation without spin
correlations fails to reproduce the correct shape of spin-sensitive observables. 
With the implementation of the correlation algorithm the decay simulation 
becomes consistent with the full (resonant) matrix element.

\subsubsection{Squark pair production in the MSSM}
\label{sec:decay_results_mssm}
To study spin correlations in a longer decay chain we now turn to the example
of squark pair production in the MSSM. We consider scalar up-quark production 
at the LHC, i.e. $pp \to \tilde{u}\tilde{u}^*$, with subsequent decays 
featuring intermediate neutralino and chargino states, i.e.

\begin{itemize}
\item $\tilde{u} \to d\,\chi_1^+ \left[ \to \chi_1^0 \, W^+ \left[ \to \mu^+ \, \nu_\mu\right]\right]$\,,
\item $\tilde{u}^* \to \bar{u}\,\chi_2^0 \left[ \to e^+\,\tilde{e}^-_R \left[\to e^-\,\chi_1^0 \right]\right]$\,.
\end{itemize}
The full final state studied thus reads 
$pp \to \tilde{u}\tilde{u}^* \to d\,\chi_1^0\,\mu^+\,\nu_\mu\;\bar{u}\,e^+\,e^-\,\chi_1^0$.
The relevant correlations are in particular the ones along the neutralino and
chargino propagators. We again compare the three different types of simulation described in 
Sec.~\ref{sec:decay_results_sm}. For our rather technical comparison we consider the
MSSM spectrum for the SPS1a benchmark point~\cite{AguilarSaavedra:2005pw}.

\begin{figure}
  \centering
  
  \begin{minipage}[t]{0.33\linewidth}
    \includegraphics[width=\textwidth]{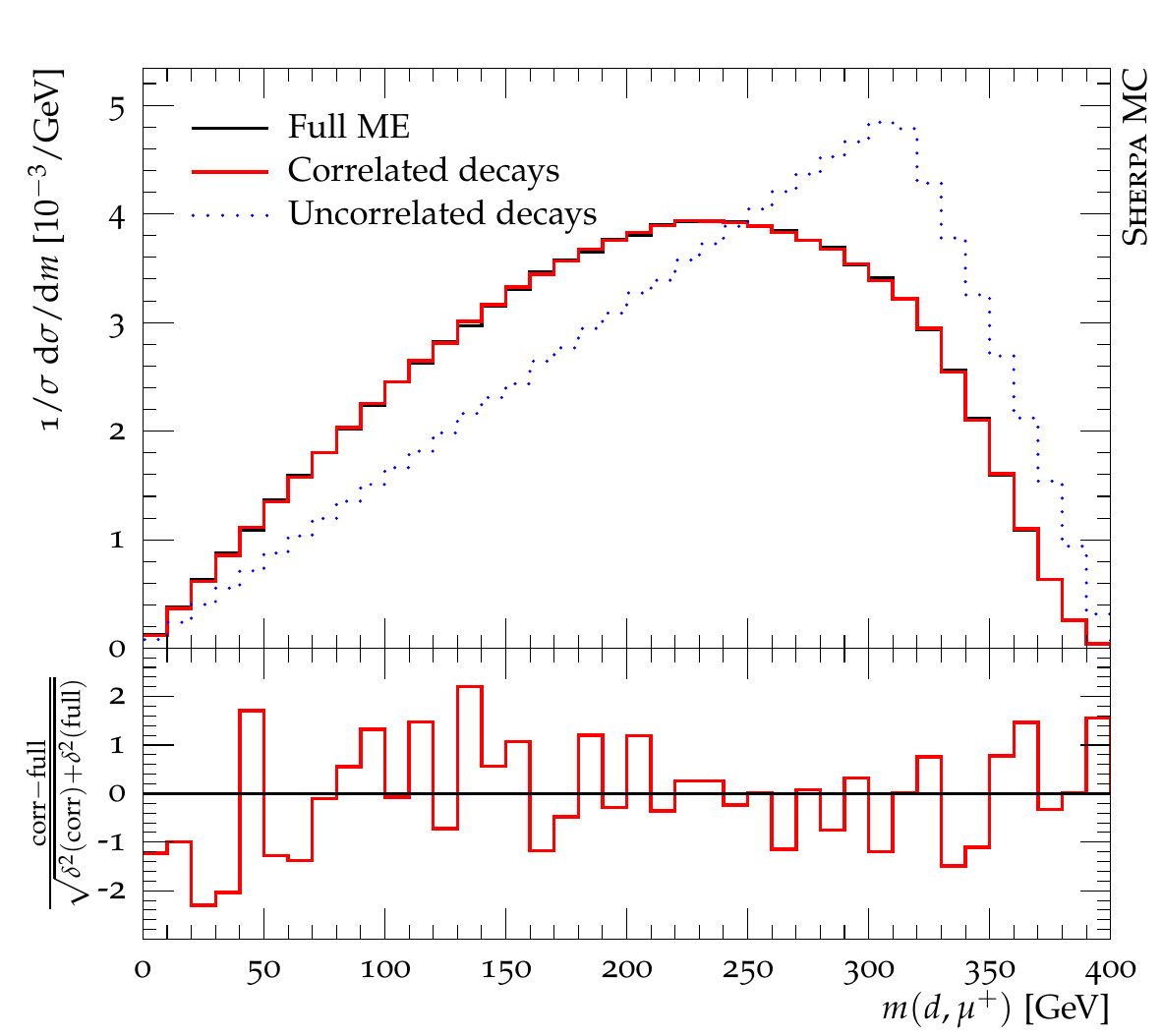}
  \end{minipage}\nolinebreak
  \begin{minipage}[t]{0.33\linewidth}
    \includegraphics[width=\textwidth]{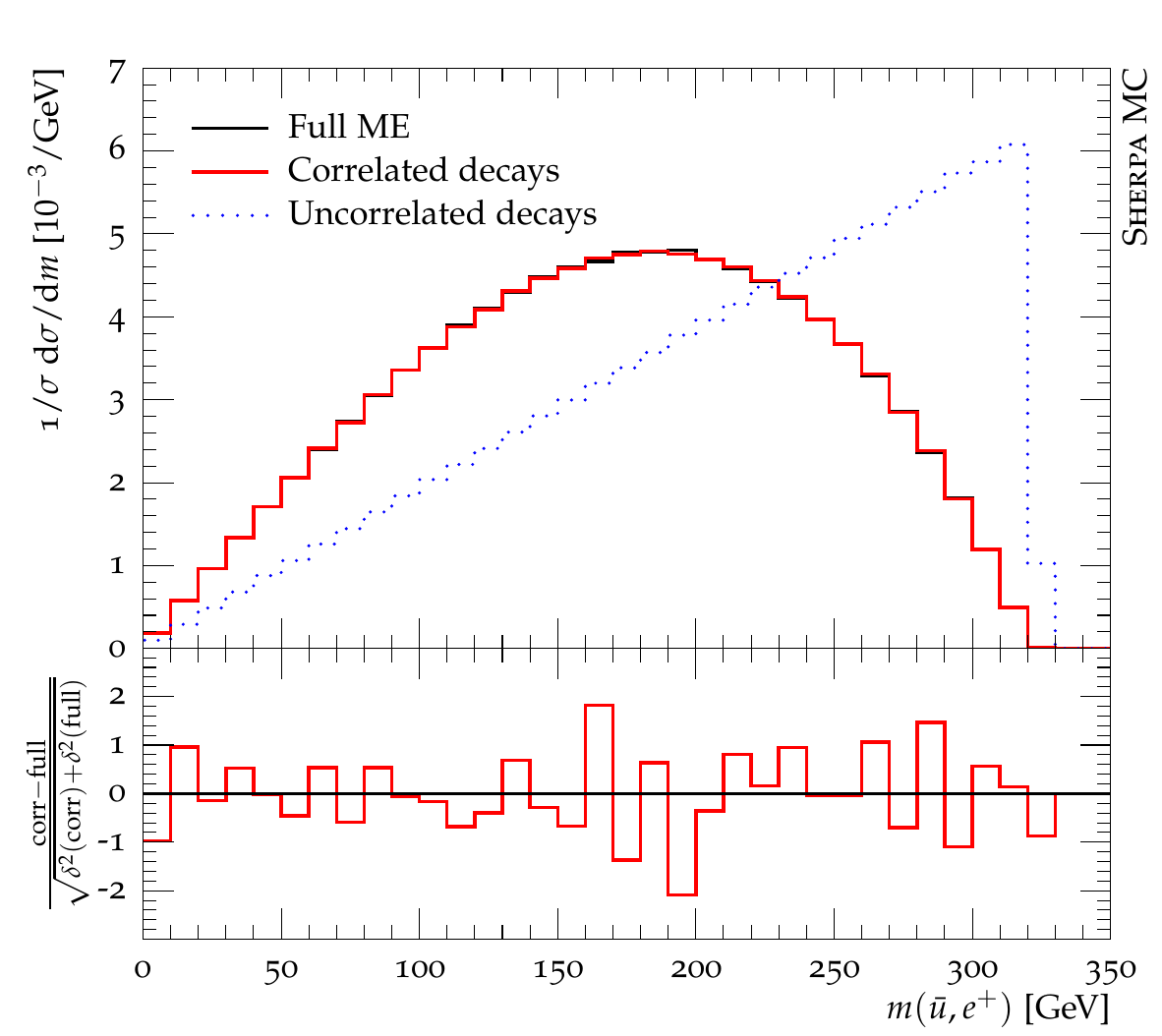}
  \end{minipage}\nolinebreak\hfill
  \begin{minipage}[t]{0.33\linewidth}
    \includegraphics[width=\textwidth]{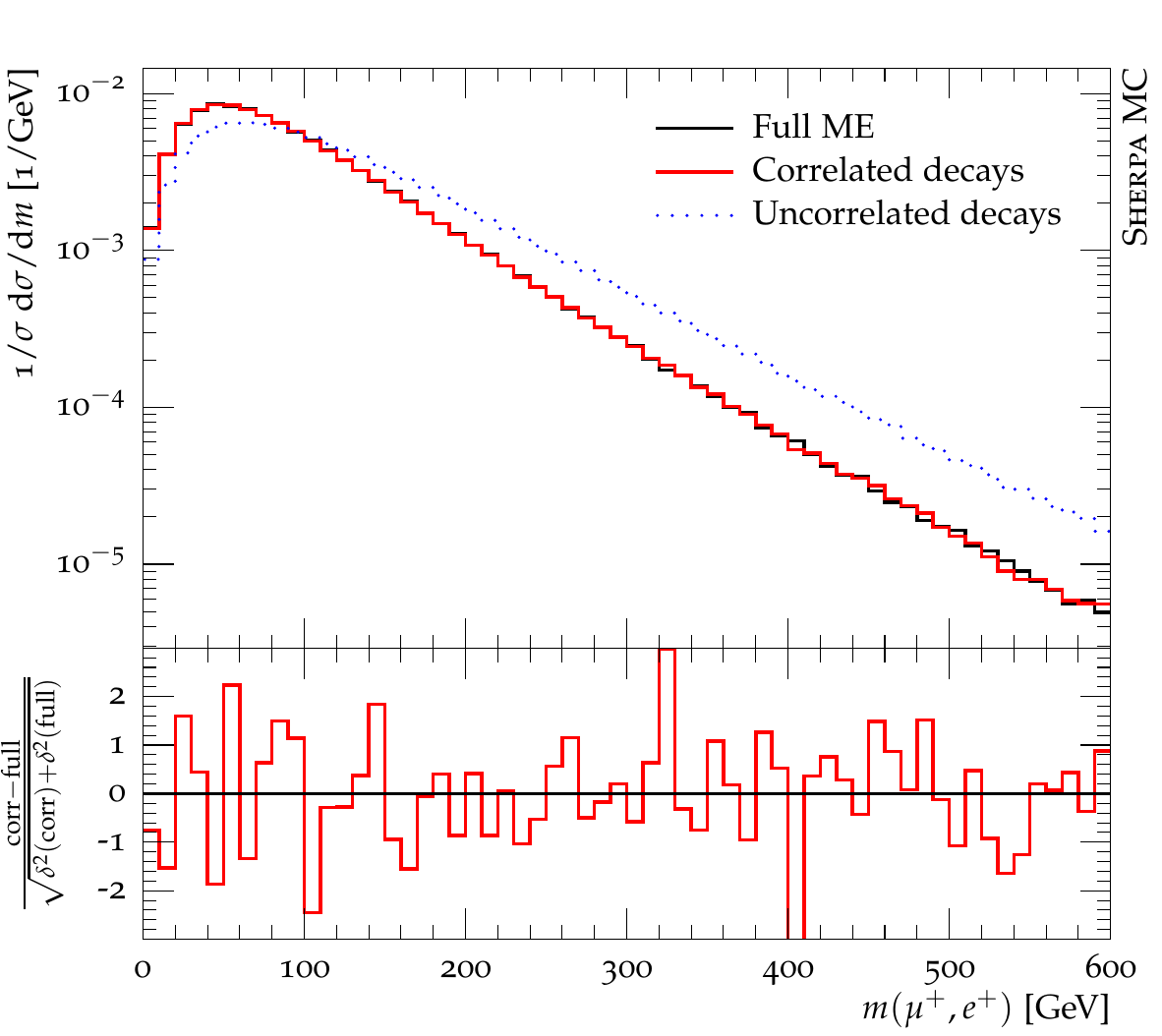}
  \end{minipage}
  \caption{Spin-correlation effects in the decay cascade following squark pair
    production in the MSSM.
    The three simulation setups are described in the main text. The ratio plot
    displays the relative difference in terms of the standard deviation and
    allows to judge the statistical compatibility between the full ME and
    correlated decay simulation.
  }
  \label{fig:decays:sc2}
\end{figure}

Fig.~\ref{fig:decays:sc2} shows three different observables which are sensitive
to spin correlations. The left panel displays the invariant mass of muon and down quark,
an observable which tests correlations within the $\tilde{u}$ decay chain. 
The middle figure shows a similar observable for the $\tilde{u}^*$ decay chain. 
The invariant mass of muon and electron displayed in the right panel demonstrates 
the impact of spin correlations in each decay chain on an observable that correlates both.

\section{Other aspects of event generation}
\label{sec:embedding}
Any simulation of new physics at the parton level must be embedded into
the full event generation at particle level in order to provide realistic
final-state information that is suitable for passing to a detector simulation
and experimental analysis. 

The combination of hard matrix elements with parton showers has been described 
in some detail in~\cite{Hoeche:2009rj,Hoeche:2009xc}.
In the context of new-physics simulations it is often necessary to amend
the merging of matrix elements and parton showers with the requirement
that no new resonances be present at higher multiplicity. This can be achieved 
in \Sherpa using a diagram filter, corresponding to the diagram-removal method
described in~\cite{Tait:1999cf,*Frixione:2008yi}.

Our simulation also includes parton-shower effects in the decay cascade. 
To account for the fact that in such a case both external and intermediate 
particles can radiate QCD quanta we use truncated showers~\cite{Schumann:2007mg,Hoeche:2009rj}
on the intermediate states. The input configuration for such a shower simulation 
is a branching history starting with the hard $2\to n$ process 
with resummation scale $\mu_Q$. For each decay process new ``layers'' 
are added to this configuration, encoding the $2\to n+1$, $2\to n+2$, \dots
final states, each with a corresponding new resummation scale for the
parton shower, that is given by the mass of the particle setting the kink
in the color flow. In the case of $t\to Wb$ decays, this would be the $W$-boson
mass, for example. 

Note that we implement parton showers in production only, not in decay. 
This means that for each decaying particle the parton shower is performed 
from the resummation scale in its production process to the particle width.
The same particle does not radiate again during its own decay, which would 
in principle be required~\cite{Hamilton:2006ms}. The mismatch resulting from 
this approximation is typically small, and we plan to include the missing effects 
in the near future. Earlier versions of Sherpa, which were based on a different 
parton shower~\cite{Krauss:2005re}, did indeed include the corresponding 
algorithm~\cite{Hoeche:2008ds,Gleisberg:2008ta}.

In addition to the QCD parton shower, \Sherpa also simulates QED emissions 
using the YFS algorithm, as detailed in~\cite{Schonherr:2008av}. This is done
before the parton shower is implemented.

Ultimately, \Sherpa invokes a cluster hadronization model~\cite{Winter:2003tt} to account for the 
fragmentation of partons into hadrons. However, our hadronization routines 
can only handle colored Standard-Model partons so far. Other long-lived or even stable 
colored particles that hadronize, as for example present in various supersymmetric 
models~\cite{Kilian:2004uj,*Barbier:2004ez}, cannot be dealt with at present.

\section{Summary and outlook}
\label{sec:conclusions}
In this publication we described the methods used to implement arbitrary 
new-physics models into the event generator \Sherpa. We provide an automatic
generator for Lorentz calculators, which allows to implement interaction vertices
which are not present in either the Standard Model or simple extensions thereof.
We also extend the matrix-element generator Comix, such that arbitrary higher-point
functions can be used for amplitude generation. The new generator supports the
Universal FeynRules Output, which is provided by programs like FeynRules and Sarah.

The new and extended version of Comix described here, together with the newly
constructed decay module of \Sherpa, allows to compute the production and decay
of new-physics particles, with spin correlations and off-shell effects in the decay
taken into account. The simulation is embedded in the larger event generation framework
of \Sherpa to also include QCD radiative corrections by means of the parton shower,
QED radiative corrections by means of the YFS approach, and non-perturbative effects 
through cluster hadronization and hadron decays. Overall, we provide a complete
framework to address many new-physics simulations in a fully automated way.
Currently our implementation is restricted to spin-0, spin-1/2 and spin-1 particles,
but the addition of higher-spin states is foreseen for the near future.

\section*{Acknowledgments}
This work was supported by the US Department of Energy under contract
DE--AC02--76SF00515. Frank Siegert's work was supported by the German
Research Foundation (DFG) under grant No.\ SI 2009/1-1. Silvan
Kuttimalai would like to thank the theory group at SLAC National 
Accelerator Laboratory for hospitality. His
work was supported by the European Union as part of the FP7 Marie 
Curie Initial Training Network MCnetITN (PITN-GA-2012-315877). Steffen 
Schumann acknowledges financial support from BMBF under contract 05H12MG5. 
Stefan H{\"o}che thanks the Center for Future High Energy Physics at IHEP for
hospitality while this work was finalized.

\bibliographystyle{bib/amsunsrt_modp}
\bibliography{bib/journal}
\end{document}